\documentclass[aps,prl,preprint,groupedaddress]{revtex4}
\usepackage{dcolumn}
\usepackage{amsmath}
\usepackage{graphicx}
\begin{document}

\preprint{BARI-TH 448/2002}

\title{New indications on the Higgs boson mass  from lattice simulations}

\author{P. Cea$^{1,2}$}
\email[]{Paolo.Cea@ba.infn.it}
\author{M. Consoli$^{3}$}
\email[]{Maurizio.Consoli@ct.infn.it}
\author{L. Cosmai$^{1}$}
\email[]{Leonardo.Cosmai@ba.infn.it}
%\homepage[]{Your web page}
%\thanks{}
%\altaffiliation{}
\affiliation{$^1$INFN - Sezione di Bari, I-70126 Bari, Italy\\
$^2$Physics Department, Univ. of Bari, I-70126 Bari, Italy\\
$^3$INFN - Sezione di Catania, I-95100 Catania, Italy}

%\date{\today}

\begin{abstract}
The `triviality' of $\Phi^4_4$ has been
 traditionally interpreted within perturbation theory where the prediction
 for the Higgs boson mass depends on the magnitude of the ultraviolet
cutoff $\Lambda$. This  approach crucially assumes that the vacuum
field and its quantum fluctuations rescale in the same way. The
results of the present lattice simulation, confirming previous
numerical indications, show that this assumption is not true. As a
consequence, large values of the Higgs mass $m_H$ can coexist with
the limit $\Lambda\to \infty $. As an example, by extrapolating to
the Standard Model our results obtained in the Ising limit of the
one-component theory, one can obtain a value as large as $m_H=760
\pm 21$ GeV, independently of $\Lambda$.
\end{abstract}


\maketitle

The `triviality' of $\Phi^4$ theories
in 3+1 space-time dimensions \cite{Sokal_book} is generally
 interpreted within perturbation theory. In this {\it interpretation},
these theories represent just an effective
description, valid only up to some cutoff scale $\Lambda$.
Without
a cutoff, the argument goes, there would be no scalar self-interactions and
without them no symmetry breaking.

This conventional view extends
to any number of scalar field components and, when used
in the Standard Model, leads to predict that
the Higgs boson mass squared,
$m^2_H$, is proportional to $g_R v^2_R$, where $v_R$ is the known weak
scale (246~GeV) and $g_R \sim 1/{\ln \Lambda}$ is the renormalized scalar
self-coupling.  Therefore, the ratio $m_H/v_R$
would be a cutoff-dependent quantity that becomes
smaller and smaller when $\Lambda$ is made larger and larger.

By accepting the validity of this
picture, there are important phenomenological implications.
For instance,
a precise measurement of $m_H$, say $m_H=760 \pm 21$ GeV,
would constrain the possible values of $\Lambda$ to be smaller
than about 2 TeV.

In an alternative approach~
\cite{Consoli:1994jr,Consoli:1997ra},
however, this conclusion is not true.
The crucial point is that
the `Higgs condensate' and its quantum fluctuations undergo different
rescalings when changing the ultraviolet cutoff. Therefore,
the relation between $m_H$ and the physical $v_R$
is not the same as in perturbation theory.

To better clarify the issue, we observe that,
beyond perturbation theory, in a
broken-symmetry phase, there are two different definitions of the field
rescaling. There is a rescaling of the `condensate', say
$Z\equiv Z_\varphi$, and a rescaling of the fluctuations, say
$Z\equiv Z_{\text{prop}}$.

To this end, let us consider a one-component scalar theory and introduce
the bare expectation value
$v_B=\langle\Phi_{\text{ latt}}\rangle$ associated with
the `lattice' field as defined at the cutoff scale.
By $Z\equiv Z_\varphi$ we mean
the rescaling that is needed to obtain
the physical vacuum field
$v_R= v_B / \sqrt{Z_\varphi}$.
By {\it physical}, we mean that the
quadratic shape of the effective potential
$V_{\text{eff}}=V_{\text{eff}}(\varphi_R)$,
evaluated at $\varphi_R=\pm v_R$, is precisely given by
$m^2_H$. Since the second derivative of the effective potential is
the zero-four-momentum two-point function,
this standard definition is equivalent to define $Z_\varphi$ as
\begin{equation}
\label{z1phi}
                          Z_\varphi= m^2_H \chi_2 (0)
\end{equation}
where $\chi_2(0)$ is the zero-momentum susceptibility.

On the other hand, $Z\equiv Z_{\text{prop}}$
is determined from the residue of the connected propagator
on its mass shell. Assuming `triviality' and
the K\'allen-Lehmann representation for the shifted quantum field,
one predicts $Z_{\text{prop}} \to 1$
when approaching the continuum theory.

Now, in the standard approach one  assumes
$Z_\varphi=Z_{\text {prop}}$ (up to small perturbative corrections).
On the other hand, in a different interpretation
of triviality~\cite{Consoli:1994jr,Consoli:1997ra}, although
$Z_{\text {prop}}\to 1$, as in leading-order perturbation theory,
$Z_\varphi\sim \ln \Lambda $ is fully non perturbative and
diverges in the continuum limit.

In this case, differently from perturbation
theory, in order
to obtain $v_R$ from the bare $v_B$ one has to apply a non-trivial
correction. As a consequence,
 $m_H$ and $v_R$ scale uniformly in the continuum
limit. From a phenomenological point of view,
 assuming to know the value of $v_R$,
a measurement of $m_H$ does not provide any information on the magnitude
of $\Lambda$ since the ratio
$C=m_H/v_R$ is a cutoff-independent quantity. Moreover,
in this approach, the quantity $C$ does not represent
the measure of any {\it observable} interaction.

The difference between $Z_\varphi$ and $Z_{\text {prop}}$
has an important physical meaning, being a
distinctive feature of the
Bose condensation phenomenon~\cite{Consoli:1999ni}.
In gaussian-like approximations to the effective potential,
one finds
$m_H/v_R=2 \pi \sqrt{2 \zeta}$, with $0< \zeta \leq 2$~\cite{Consoli:1999ni},
$\zeta$
being a cutoff-independent number determined by the quadratic shape of
the effective potential
$V_{\text {eff}}(\varphi_R)$
at $\varphi_R=0$. For instance, $\zeta=1$
corresponds to the classically scale-invariant case or `Coleman-Weinberg
regime'.

To check the alternative picture of Refs.
\cite{Consoli:1994jr,Consoli:1997ra} against the generally accepted point of
view, one can run numerical simulations of the theory. In this respect,
we observe that
numerical evidence for different cutoff dependencies of
$Z_\varphi$ and $ Z_{\text{prop}}$ has already been reported in
Refs.~\cite{Cea:1998hy,Cea:1999kn,Cea:1999zu}. In those calculations,
performed in the Ising limit of the one-component theory,
one was fitting the lattice data
for the connected propagator to the (lattice version of the)
two-parameter form
\begin{equation}
\label{gprop}
           G_{\text{fit}}(p)= \frac{Z_{\text{prop}}}{ p^2 + m^2_{\text{latt}} }
\end{equation}
After computing the zero-momentum susceptibility $\chi_{\text{latt}}$,
it was possible to compare the value of
$Z_\varphi \equiv m^2_{\text{latt}} \chi_{\text{latt}}$
 with the fitted $Z_{\text{prop}}$,
both in the symmetric and broken phases.
While no difference was found in the
symmetric phase, $Z_\varphi$ and $Z_{\text{prop}}$ were found to be
 sizeably different
in the broken phase. In particular, $Z_{\text{prop}}$ was very slowly
varying and steadily approaching unity from below in the continuum limit.
$Z_{\varphi}$, on the other hand,
was found to rapidly increase {\it above} unity in the same limit.

A possible objection to this strategy is that the two-parameter form
Eq.(\ref{gprop}), although providing a good description of
the lattice data, neglects higher-order corrections to the structure
of the propagator. As a consequence, one might object that the
extraction of the various parameters
is affected in an uncontrolled way (even though the fitted $Z_{\text{prop}}$
was found~\cite{Cea:1998hy,Cea:1999kn}
in good agreement with its perturbative prediction).

For this reason, we have decided to change strategy by
performing a new set of lattice calculations.
Rather than studying the propagator,
we have addressed the model-independent  lattice
measurement of
the susceptibility. In this way, {\it assuming} the
mass values from
perturbation theory, one can obtain a precise
determination of $Z_\varphi$ that can be compared
with the perturbative predictions.
Our results, will be presented in the following.

For our simulations we have considered again
the Ising limit of a one-component $\Phi^4_4$ theory. Traditionally, this
has been considered as a convenient laboratory to obtain non-perturbative
information on the theory and corresponds to
the lattice action
\begin{equation}
\label{ising}
S_{\rm ising}= - \kappa \sum _x \sum _\mu [
\phi(x + \hat{e}_\mu) \phi(x) +\phi(x - \hat{e}_\mu) \phi(x)]
\end{equation}
where $\phi(x)=\pm 1$. In an infinite lattice, the broken phase
is found for $\kappa > 0.07475$.

We performed Monte-Carlo simulations of this Ising action using the
Swendsen-Wang~\cite{Swendsen:1987ce} and Wolff~\cite{Wolff:1989uh} cluster algorithms to compute
the zero-momentum susceptibility
\begin{equation}
\label{chi}
   \chi_{\text{latt}} = L^4 [ \langle |\phi|^2 \rangle - \langle|\phi|\rangle^2]
\end{equation}
As a check of the validity of our algorithms, we show in Table~\ref{tab:table0} a
comparison with previous determinations of $\chi_{\text {latt}}$
obtained by other authors.
\begin{table}[t]
\caption{\label{tab:table0}
We compare our determinations of $\langle |\phi| \rangle$ and $\chi_{\text{latt}}$ for given $\kappa$
with corresponding determinations found in the literature (Ref.~\cite{Jansen:1989cw}).
In the algorithm column, 'S-W' stands for the Swendsen-Wang algorithm~\cite{Swendsen:1987ce},
while 'W' stands for the Wolff algorithm~\cite{Wolff:1989uh}. 'Ksweeps' stands for
sweeps multiplied by $10^3$.}
\begin{ruledtabular}
\begin{tabular}{cccccc}
$\kappa$   &lattice  &algorithm &Ksweeps
& $\langle |\phi| \rangle$  &$\chi_{\text{latt}}$  \\
\hline
0.077  &$32^4$  &S-W  & 3500  & 0.38951(1) &18.21(4)   \\
0.077  &$16^4$  &Ref.~\cite{Jansen:1989cw}  & 10000 & 0.38947(2)  &18.18(2)   \\ \hline
0.076  &$20^4$  &W  & 400  & 0.30165(8) &37.59(31)   \\
0.076  &$20^4$  &Ref.~\cite{Jansen:1989cw}  & 7500  & 0.30158(2) &37.85(6)   \\
\end{tabular}
\end{ruledtabular}
\end{table}

To compare our results with perturbation theory, we have adopted
the L\"uscher-Weisz scheme~\cite{Luscher:1988ek} where the prediction for the ratio
$m_H/v_R$ can be expressed as
\begin{equation}
\label{gR}
             \left[ \frac{m_H}{v_R} \right]_{\text{LW}} \equiv \sqrt  {  \frac{g_R}{3} }
\end{equation}
Assuming the values of $g_R$ reported in the second column of Table~3
of Ref.~\cite{Luscher:1988ek}, the ratio in Eq.(\ref{gR}) becomes smaller and smaller
when approaching the continuum limit.

As anticipated, to check the consistency of this prediction,
we shall adopt the perturbative input values for the mass
and denote
by $m_{\text{input}}$ the value of the parameter $m_R$ reported in the
first column of Table~3 in Ref.~\cite{Luscher:1988ek} for any value of $\kappa$
(the Ising limit corresponding
to the value of the other parameter $\bar{\lambda}=1$). In this way,
computing the susceptibility on the lattice,
we shall compare the quantity
\begin{equation}
\label{zphi}
       Z_\varphi\equiv 2\kappa m^2_{\text{input}} \chi_{\text{latt}}
\end{equation}
with the perturbative prediction for
$Z_{\text{LW}}\equiv 2\kappa Z_R$
where $Z_R$ is defined in the third
column of Table~3 in Ref.~\cite{Luscher:1988ek}.
\begin{table}[t]
\caption{\label{tab:table1}
The details of the lattice simulations for each $\kappa$ corresponding to $m_{\text{input}}$.
In the algorithm column, 'S-W' stands for the Swendsen-Wang algorithm~\cite{Swendsen:1987ce},
while 'W' stands for the Wolff algorithm~\cite{Wolff:1989uh}. 'Ksweeps' stands for
sweeps multiplied by $10^3$.}
\begin{ruledtabular}
\begin{tabular}{cccccc}
$m_{\text{input}}$   &$\kappa$   &lattice  &algorithm &Ksweeps
&$\chi_{\text{latt}}$  \\
\hline
0.4  &0.0759  &$32^4$  &S-W  & 1750  &41.714 (0.132)  \\
0.4  &0.0759  &$48^4$  &W    &   60  &41.948 (0.927)  \\ \hline
0.3  &0.0754  &$32^4$  &S-W  &  345  &87.449 (0.758)  \\
0.3  &0.0754  &$48^4$  &W    &  406  &87.821 (0.555)  \\ \hline
0.2  &0.0751  &$48^4$  &W    &   27  &203.828 (3.058)  \\
0.2  &0.0751  &$52^4$  &W    &   48  &201.191 (6.140)  \\
0.2  &0.0751  &$60^4$  &W    &    7  &202.398 (8.614)  \\ \hline
0.1  &0.0749  &$68^4$  &W    &   24  &1125.444 (36.365)  \\
0.1  &0.0749  &$72^4$  &W    &    8  &1140.880 (39.025)  \\
\end{tabular}
\end{ruledtabular}
\end{table}

Our lattice results for $\chi_{\text{latt}}$ are reported
in Table~\ref{tab:table1} for the different values of $\kappa$ corresponding to
$m_{\text{input}}=0.4, 0.3, 0.2, 0.1$.
In Table~~\ref{tab:table1} we have also indicated the algorithm used
for upgrading the lattice configurations and
the number of sweeps at each value of $\kappa$ and lattice size. In the case of the Wolff algorithm
the number of sweeps is the number of Wolff sweeps multiplied by the ratio between the average
cluster size and the lattice volume.  We used different lattice sizes at each value of $\kappa$
to have a check of the finite-size effects.
The statistical errors have been estimated using the jackknife.
\begin{table}[t]
\caption{\label{tab:table2}
The values of $g_R$ and $Z_{\text{LW}}$ for each  $m_{\text{input}}$ as given in Table~3
of Ref.~\cite{Luscher:1988ek}. $Z_\varphi$ is defined in Eq.~(\ref{zphi}). The errors quoted on
$Z_\varphi$ are only due to the statistical uncertainty  of $\chi_{\text{latt}}$
(see Table~\ref{tab:table1}).}
\begin{ruledtabular}
\begin{tabular}{cccc}
$m_{\text{input}}$   &$g_R$  &$Z_{\text{LW}}$    &$Z_\varphi$    \\
\hline
0.4  &27 (2) &0.929 (14)   &1.019 (23)  \\
0.3  &24 (2) &0.932 (14)   &1.192 (8)  \\
0.2  &20 (1) &0.938 (12)   &1.216 (52)  \\
0.1  &16.4 (9) &0.944 (11)   &1.709 (58)  \\
\end{tabular}
\end{ruledtabular}
\end{table}

We have reported in Table~\ref{tab:table2}, the corresponding entries
for $Z_\varphi$, $Z_{\text{LW}}$ and $g_R$.
As one can see, the two Z's are
sizeably different and the discrepancy becomes larger and larger
 when approaching the continuum limit, precisely the same trend found in
Refs.\cite{Cea:1998hy,Cea:1999kn}. This confirms that, approaching the continuum limit,
the rescaling of the `Higgs condensate'
cannot be described in perturbation theory.

Now, if zero-momentum quantities rescale differently from the
perturbative predictions, one may wonder about the
relation between $m_H$ and $v_R$, when this
is rescaled through $Z \equiv Z_\varphi$
rather than through the perturbative $Z\equiv Z_{\text{LW}}$.
 In this case, one finds the alternative relation
\begin{equation}
\label{mh}
\frac{m_H}{v_R}= \sqrt{ \frac{g_R}{3} \frac{Z_\varphi}{Z_{\text{LW}} } }
\equiv C
\end{equation}
obtained by replacing
$Z_{\text{LW}} \to Z_\varphi$ in
Ref.~\cite{Luscher:1988ek}  but correcting for the perturbative $Z_{\text{LW}}$ introduced
in the L\"uscher and Weisz approach.
\begin{figure}[t]
\includegraphics[width=0.8\textwidth,clip]{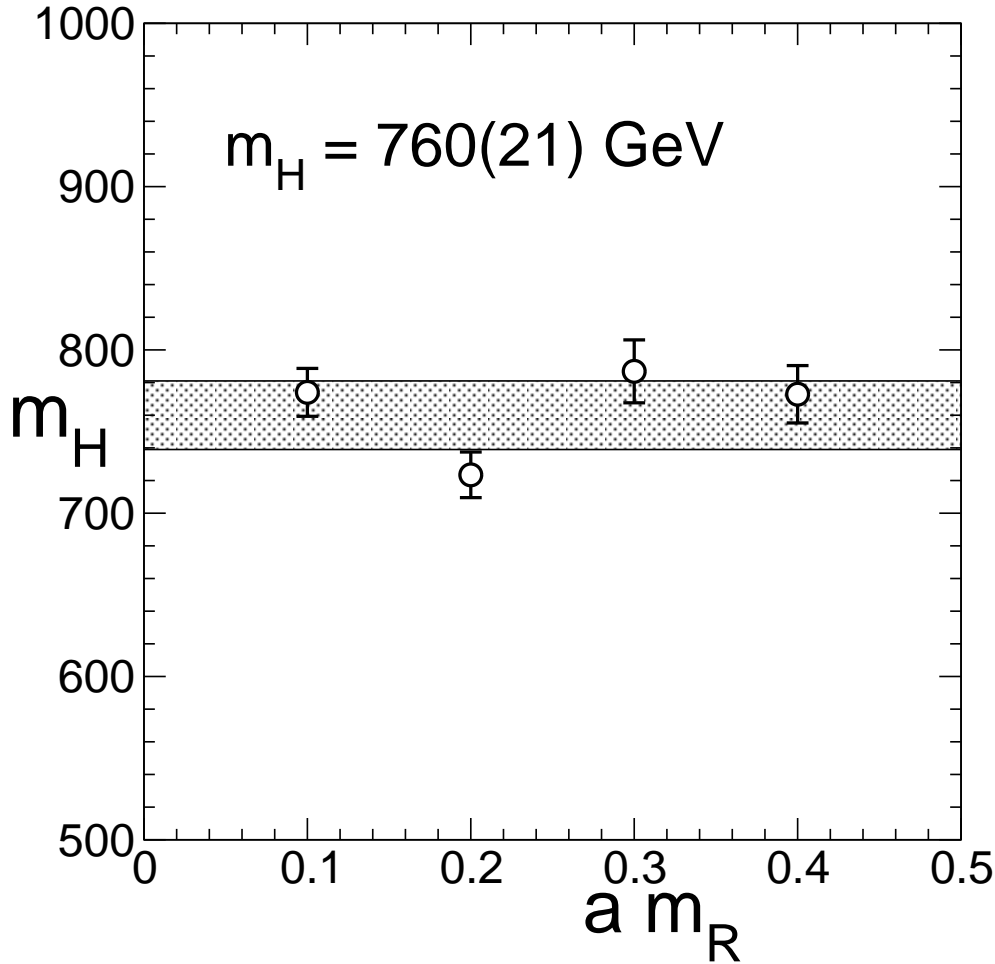}
\caption{\label{fig:01} The values of $m_H$ as defined through Eq.~(\ref{mh}) versus
$m_{\text{input}}=a m_R$. The error band corresponds to a one standard deviation
error in the determination of $m_H$ through a fit with a constant function.}
\end{figure}

According to the picture of Refs.~\cite{Consoli:1994jr,Consoli:1997ra}, one expects
$Z_\varphi \sim \ln \Lambda$ to compensate the $1/\ln \Lambda$ from $g_R$
so that $C$ should be a cutoff-independent constant. To this end, one
can check the values of
$Z_\varphi$, $Z_{\text{LW}}$ and $g_R$ in our Table~\ref{tab:table2}.
We find that $C$
is a constant, to a good approximation,
$C=3.087 \pm 0.084$. As an example, this value, when
combined with the Standard
Model value $v_R = 246$ GeV, would yield a Higgs mass
$m_H= 760 \pm 21$~GeV independently of the ultraviolet cutoff
$\Lambda \sim \pi/a$ (see Fig.~1).

Notice that this value is {\it not} a
prediction for the mass of the Higgs boson,
neither in the one-component theory
nor in the Standard Model. In fact, the uncertainty
is dominated by the statistical error in $\chi_{\rm latt}$
at any value of $\kappa$ and neglects any
{\it theoretical} uncertainty associated with
approaching the critical line in the Ising limit.
Traditionally, the Ising limit corresponds to the maximal value of
$m_H/v_R$, as determined from the perturbative trend Eq.~(\ref{gR})
with $g_R/3 \sim A/ \ln \Lambda$. However, our simulation show that
one is faced with the more general scenario Eq.~(\ref{mh}) where
$Z_\varphi \sim B \ln \Lambda$ so that
$m_H/v_R \sim \sqrt{AB}$.

In this sense, the implications of our
results for the Standard Model are mainly
of `qualitative' nature and amount
to the statement that the value of the Higgs boson mass, in units of
246 GeV, does not depend on the magnitude of the ultraviolet cutoff.
However, as a consequence of our results,
the whole issue of the upper bounds on the Higgs mass
 is affected suggesting the need of more
extensive studies of the critical line to compare the possible values
of $C= \sqrt{AB}$ with the value
$C_{\text{Ising}} \simeq \pi$ obtained in the Ising limit.
This should also
be performed in the O(4)-symmetric case which, after all,
is the one relevant for the Standard Model.
Independently of this more refined analysis, it is also true
that a value as large as $m_H=760 \pm 21$~GeV,
would be in good agreement with
a recent phenomenological
analysis of  radiative corrections~\cite{Loinaz:2002ep} that
points toward substantially
larger Higgs masses than previously obtained through
global fits to Standard Model observables.

%\bibliography{phi4}

\end{document}